\documentclass[prd,onecolumn,nofootinbib,aps,tightenlines,preprintnumbers,notitlepage,longbibliography,superscriptaddress,12pt]{revtex4-1}

\usepackage{multirow}
\usepackage{hyperref}
\usepackage{graphicx}
\usepackage{bm}
\usepackage{color, colortbl, xcolor}
\usepackage{subcaption} % For subfigures
\usepackage{amsmath}
\usepackage{amssymb}
\usepackage{wasysym}
\usepackage{ulem}

\newcommand{\be}{\begin{eqnarray}}
\newcommand{\ee}{\end{eqnarray}}

\newcommand{\beq}{\begin{equation}}
\newcommand{\eeq}{\end{equation}}

\newcommand{\caps}{Center for AstroPhysical Surveys, National Center for Supercomputing Applications, University of Illinois Urbana-Champaign, Urbana, IL, 61801, USA}

\newcommand{\iap}{Institut d'Astrophysique de Paris, CNRS UMR 7095, 98 bis Bd Arago, 75014 Paris, France.}

\begin{document}

\title{The End of Easy Phenomenology for CMB Experiments: A Case Study in the Dark Sector}

\author{Cynthia~Trendafilova}
\email[]{ctrendaf@illinois.edu}
\affiliation{\caps}

\author{Ali~Rida~Khalife}
\email[]{ridakhal@iap.fr}
\affiliation{\iap}

\author{Silvia~Galli}
\email[]{gallis@iap.fr}
\affiliation{\iap}

\date{\today}

\begin{abstract}
The precision of Cosmic Microwave Background (CMB) experiments, specifically its lensing reconstruction, has reached the limit where non-linear corrections cannot be ignored. Neglecting these corrections results in biased constraints on cosmological parameters. In this work, we use lensing data from {\it{Planck}} and the South Pole Telescope third generation camera (SPT-3G) taken in 2018 to highlight the impact of these biases in two ways. First, we estimate the shifts due to ignoring non-linear corrections in $\Lambda$CDM. We find 0.2-0.6$\sigma$ shifts in the Dark Matter (DM) fraction, the Hubble constant, and the amplitude of matter fluctuations at $8 h^{-1}$ Mpc. Second, we estimate the loss in constraining power for not including data sensitive to non-linear corrections. As a case study, we consider two interacting DM models, for which such corrections are not readily available in existing CMB Boltzmann codes. The first one is DM interacting with baryons, while the second is DM interacting with Dark Radiation (DR). 
For the former case, when we add primary CMB data from SPT-3G 2018 observations, we find that constraints on model parameters improve by 10-20\% compared to previous studies.
However, we forecast a further 50\% improvement on these constraints if one could include current or upcoming SPT-3G lensing data. For the case of DM interacting with DR, no meaningful constraints on the model parameters are found without including information from CMB lensing. We also highlight that the codes used to calculate non-linear corrections in $\Lambda$CDM, specifically \texttt{HaloFit} and \texttt{HMCode}, provide unstable results when improperly used for these extended models.
These outcomes constitute a reason for caution if using CMB lensing data when constraining such models, as well as a motivation for estimating their non-linear corrections.
\end{abstract}

\maketitle

\section{Introduction}
\label{sec:Introduction}
Providing the earliest visible snapshot of our universe, the CMB has become a key tool for learning about the fundamental physical properties of our universe~\cite{Penzias_Wilson,Planck:2015,Planck:2018}. By carrying information from the time of recombination, the CMB serves as a laboratory for physical conditions which are currently inaccessible to us in terrestrial laboratory experiments.
The properties of the CMB and large-scale structure (LSS) data are well-described by the standard six-parameter $\Lambda$CDM model, whose parameters have been measured with great accuracy by CMB experiments, including the {\it{Planck}} satellite~\cite{Planck:2015,Planck:2018}, the Atacama Cosmology Telescope (ACT)~\cite{ACT_DR4,ACT:2023kun}, and the South Pole Telescope (SPT)~\cite{Dutcher_etal,Balkenhol_etal}. 
The satellite-based measurements of the CMB have been complemented by ground-based measurements aiming to more precisely probe the small-scale temperature and polarization anisotropies. 
The anisotropies of the CMB will be further probed by upcoming measurements from surveys such as the SPT-3G+~\cite{SPT3Gplus}, Simons Observatory~\cite{SimonsObservatory:2018koc}, and CMB-S4~\cite{CMB-S4:2016ple}. 

The contemporary CMB datasets used in current analyses have probed structure formation either directly through the reconstructed lensing power or through its impact on the lensed CMB temperature and polarization power spectra. However, these datasets probe scales which are not strongly sensitive to the physics of structure growth on small scales where non-linear effects become significant. With newer lensing datasets, this is changing; recent polarization-based lensing reconstruction from SPT data~\cite{SPT-3G:2024atg} has detected a non-zero amplitude of non-linear structure growth at $> 3\sigma$, the first time from CMB lensing data.
Future CMB experiments will probe this at greater sensitivity, in addition to probing the damping tail of the primary CMB power spectra. 
Because we do not observe the unlensed CMB directly but rather the lensed CMB, these measurements will also be sensitive to the details of the modeling of the LSS growth~\cite{McCarthy:2021lfp}.
Therefore, such details will become an increasing source of systematic errors for upcoming CMB datasets if this physics is not modeled accurately and robustly.
Although there have been significant efforts in developing the so-called halo model~\cite{Halo_model1,Halo_model2}, used to compute theoretical predictions of CMB lensing power and lensed power spectra, this requires significant computational efforts via N-body simulations of structure growth. It is not trivial to extend such efforts to encapsulate the wide landscape of Beyond-$\Lambda$CDM models of interest to the physics community, although this has been done for a few extensions (see for example~\cite{HaloFit_Mnu,Non_Lin_Beyond_LCDM1,Non_Lin_Beyond_LCDM2,Non_Lin_Beyond_LCDM3}). When CMB datasets are applied to search for signs of these physical interactions, one must be mindful of the limitations of the theoretical tools which are being used to compute cosmological predictions.
A particular example of these limitations that quantifies and illustrates their relevance for CMB analyses can be found in the dark sector that can interact with the visible one.

Although the $\Lambda$CDM model is very successful, there are observations which remain unexplained, such as the $H_0$ and $\sigma_8$ tensions~\cite{Planck:2013,Verde_Reiss_Tension,Riess_Tension,KiDS-450}. These are accompanied by open questions in the Standard Model of Particle Physics, such as the presence of DM, the observed baryon asymmetry, and the flavor puzzle~\cite{Flavor_puzzle1,Flavor_puzzle2}. In proposing models of DM to explain our gravitational observations, it has thus been natural to leverage these models to address these other questions as well. This has led to a vast landscape of rich DM models with varied phenomenology, such as having multiple species of particles in the dark sector and interactions with the dark or visible sectors~\cite{ParticleDataGroup:2024cfk}.
In particular, two classes of dark sector interactions, which can be present in the early and late-time universe, have their respective imprints on the temperature, polarization, and lensing power spectra of the CMB. These are the scattering of DM with baryons (DM-b), and the interaction of DM with DR (DM-DR), both of which are implemented in the Boltzmann code \texttt{CLASS}~\cite{Blas:2011rf,CLASS1}. Phenomenologically, the impact of these models on the CMB can be parametrized with a few parameters, allowing one to place constraints in a way that is independent of the details of the underlying particle physics.
Previous works have applied various CMB datasets, including {\it{Planck}} primary CMB and lensing measurements~\cite{Becker:2020hzj,Lesgourgues:2015wza,Archidiacono:2019wdp,Brinckmann:2022ajr,SPT_DES_DMDR}, and ACT primary anisotropy measurements~\cite{Li:2022mdj}, to probe these models.

In this work, we study these two dark sector models in the context of searching for signals of new physics when modeling of the non-linear structure formation is lacking. We quantify the shifts in $\Lambda$CDM parameter values with current CMB and lensing datasets when parameter fitting is performed without the inclusion of non-linear corrections. We then restrict ourselves to using CMB datasets where these shifts are not significant, and we compute new constraints on these models.
We include the addition of the SPT-3G 2018 TT/TE/EE~\cite{Balkenhol_etal} data which has not been previously applied in these analyses, along with newly including ACT DR-4 TT/TE/EE~\cite{ACT_DR4} data for DM-DR interactions. 
The unbiased constraints that we report are those that do not include lensing data from any of the considered CMB experiments. As mentioned before, CMB lensing is sensitive to physics at non-linear scales, which produces biases in parameter constraints when modeling for the non-linear matter power is lacking. Nonetheless, to show the valuable constraining power of CMB lensing for these models, we present an example including this data and show the improved constraints. We stress, however, that the constraints resulting from this run are not to be used exactly, as they include the biases previously mentioned.

Our paper is structured as follows. 
In Section~\ref{Sec:Data_Sets}, we list the cosmological datasets that we consider for our analysis and define our notation for combining datasets.
In Section~\ref{Sec:Nonlinear}, we evaluate the cosmological $\Lambda$CDM parameter biases that would be incurred by using various CMB datasets without non-linear matter power spectrum corrections from the theory code.
The dark sector models which we choose to study, in order to provide a concrete example of these considerations with regards to specific models, are described in Section~\ref{sec:Models}. In Section~\ref{sec:Pk}, we illustrate our theoretical limitations by presenting the outputs of non-linear matter power spectrum codes which have not been calibrated to account for these models.
In Section~\ref{Sec:Results}, we present our results on DM-b and DM-DR model parameter constraints using the application of datasets which we consider safe to use. Finally, we conclude in Section~\ref{Sec:Conclusion}.

\section{Data Sets}
\label{Sec:Data_Sets}
Before delving into more details, we present the main data sets that we consider in our analysis, which are\footnote{For CMB experiments, for brevity, we only cite the maximum multipole of each data set since it is the most relevant for non-linear corrections.}:
\begin{enumerate}
    \item Planck: Primary CMB power spectra from {\it{Planck}} 2018 for TT, TE, EE +lowE~\cite{Planck:2018}, with multipole range $\ell\leq2508$ for TT, $\ell\leq1996$ for TE and EE.
    \item Plancklens: CMB lensing power spectrum from {\it{Planck}} 2018~\cite{Planck_Lensing}, with multipole range $L\leq400$.
    \item SPT: Primary CMB power spectra from SPT-3G 2018 for TT, TE, and EE~\cite{Balkenhol_etal,Dutcher_etal} in the multipole range $\ell< 3000$.
    \item SPTlens: Lensing potential power spectrum from SPT-3G 2018~\cite{SPT2018lens} in the multipole range $L<2000$, reconstructed from $TT$ data only.
    \item ACT: Primary CMB power spectrum data from the Atacama Cosmology Telescope DR4 for TT, TE, and EE~\cite{ACT_DR4,ACT1}. We consider data points only up to $\ell\leq3000$.\footnote{When combined with Planck, we exclude data at $\ell<1800$ in TT, as recommended in~\cite{ACT_DR4}.}
    \item ACTlens: Lensing potential power spectrum from ACT DR6~\cite{ACT:2023dou,ACT:2023kun} in the multipole range $L<1000$. 
    \item BAO: BAO measurements from the Sloan Digital Sky Survey (SDSS)~\cite{SDSS_6dF,SDSS_DR12,SDSS_DR7} and the Dark Energy Spectroscopic Instrument (DESI)~\cite{DESI_ER}, as described in Section 3.3 of~\cite{DESI_main}. We include only the BAO feature and exclude any data at non-linear scales.
\end{enumerate}
For the DM-DR runs, we use SPT, SPTlens, ACT and ACTlens in our analysis as implemented in the likelihood framework \texttt{candl}\footnote{\url{https://github.com/Lbalkenhol/candl}}~\cite{candl}.
Moreover, for ease of notation, we use the following abbreviations for different data combinations:
\begin{itemize}
    \item PB: Planck + BAO.
    \item PBS: Planck + BAO + SPT.
    \item PBSA: Planck + BAO + SPT + ACT.
    \item PBSLAL: Planck + Plancklens + BAO + SPT + SPTlens + ACT + ACTlens.
\end{itemize}

\section{$\Lambda$CDM: Non-linear corrections with Lensing data}
\label{Sec:Nonlinear}
Non-linear corrections to the matter power spectrum in $\Lambda$CDM\footnote{Non-linear corrections when massive neutrinos are included have been computed in~\cite{HaloFit_Mnu}.} are well-known and numerically implemented in codes such as \texttt{HaloFit}~\cite{HaloFit,HaloFit_Mnu,Takahashi:2012em} and \texttt{HMCode}~\cite{HMcode}. As mentioned in the Introduction~\ref{sec:Introduction}, previous CMB experiments did not have enough signal-to-noise (S/N) at the non-linear scales, which means that the presence of these corrections have a negligible impact on cosmological constraints. In this section, we perform a simple test of the previous statement for {\it{Planck}} and SPT-3G data, in the context of $\Lambda$CDM. 
Because we do not have non-linear prescriptions accounting for dark sector interactions on the scales of interest, we can at best test for biases in $\Lambda$CDM parameters only. The potential biases in our extended model parameters are unknown, however performing full N-body simulations to quantify non-linear effects for the models considered here is beyond the scope of this work, and we leave these implementations to future work.

Starting with primary CMB data from each experiment, we run a Monte Carlo Markov Chain (MCMC) with non-linear corrections from \texttt{HaloFit} included and compare the resultant posteriors to the case when non-linear corrections are excluded.
We note that even CMB temperature and polarization data may be sensitive to non-linear physics, since the lensed CMB power depends on the CMB lensing power spectrum, and therefore we test these datasets in addition to the lensing data.
We free only the six $\Lambda$CDM parameters, which can be parameterized for example by 
\be
p\in\{\Omega_bh^2,\Omega_ch^2,H_0,A_s,\tau,n_s\}.
\label{eq:LCDM_params}
\ee
These parameters are the physical density of baryons and cold DM, the Hubble constant (with its reduced form $h=H_0/100$ Mpc s km$^{-1}$), the amplitude of scalar perturbations, the optical depth to reionization, and the spectral index of scalar perturbations, respectively.
Our MCMC runs are done with \texttt{COBAYA}~\cite{COBAYA1,COBAYA2,COBAYA3}, using \texttt{CLASS}~\cite{2011JCAP...07..034B,CLASS1} as the theory code. We compare these two runs by computing
\beq
D = \left( \sum_{i,j} \left(\hat{p}_{i,\mathrm{NL}} - \hat{p}_{i,\mathrm{L}}\right) \left[\mathrm{Cov}_{\mathrm{NL}}\right]^{-1}_{ij} \left(\hat{p}_{j,\mathrm{NL}} - \hat{p}_{j,\mathrm{L}}\right) \right)^{1/2}
\label{eq:biasN}
\eeq
for some set of parameter(s) $p_i$, where $\mathrm{Cov}_{\mathrm{NL}}$ is the covariance matrix of the parameters $p_i$ from the run where non-linear corrections are included. Note that this expression reduces to 
\beq
\frac{\Delta p}{\sigma} = \frac{|\hat{p}_\mathrm{NL} - \hat{p}_\mathrm{L}|}{{\sigma_\mathrm{NL}}}
\label{eq:bias1}
\eeq
in the case of a single parameter.
In addition to the parameters defined above, we also consider the derived parameter $S_8 = \sigma_8\sqrt{\Omega_m/0.3}$, where $\sigma_8$ is the amplitude of matter fluctuations at $8h^{-1}$Mpc and $\Omega_m$ is the matter density parameter. Moreover, $\hat{p}$ corresponds to the mean of $p$, the subscript NL (L) corresponds to the case where non-linear corrections are included (excluded) and $\sigma$ is the standard deviation. 
What we find for the case of primary CMB data only is at most a $\sim0.15\sigma$ shift in parameters, which means these data from these two experiments are marginally sensitive to non-linear corrections.

\begin{figure}[h!]
    \centering
    \begin{subfigure}{0.45\textwidth}
        \includegraphics[width=\textwidth]{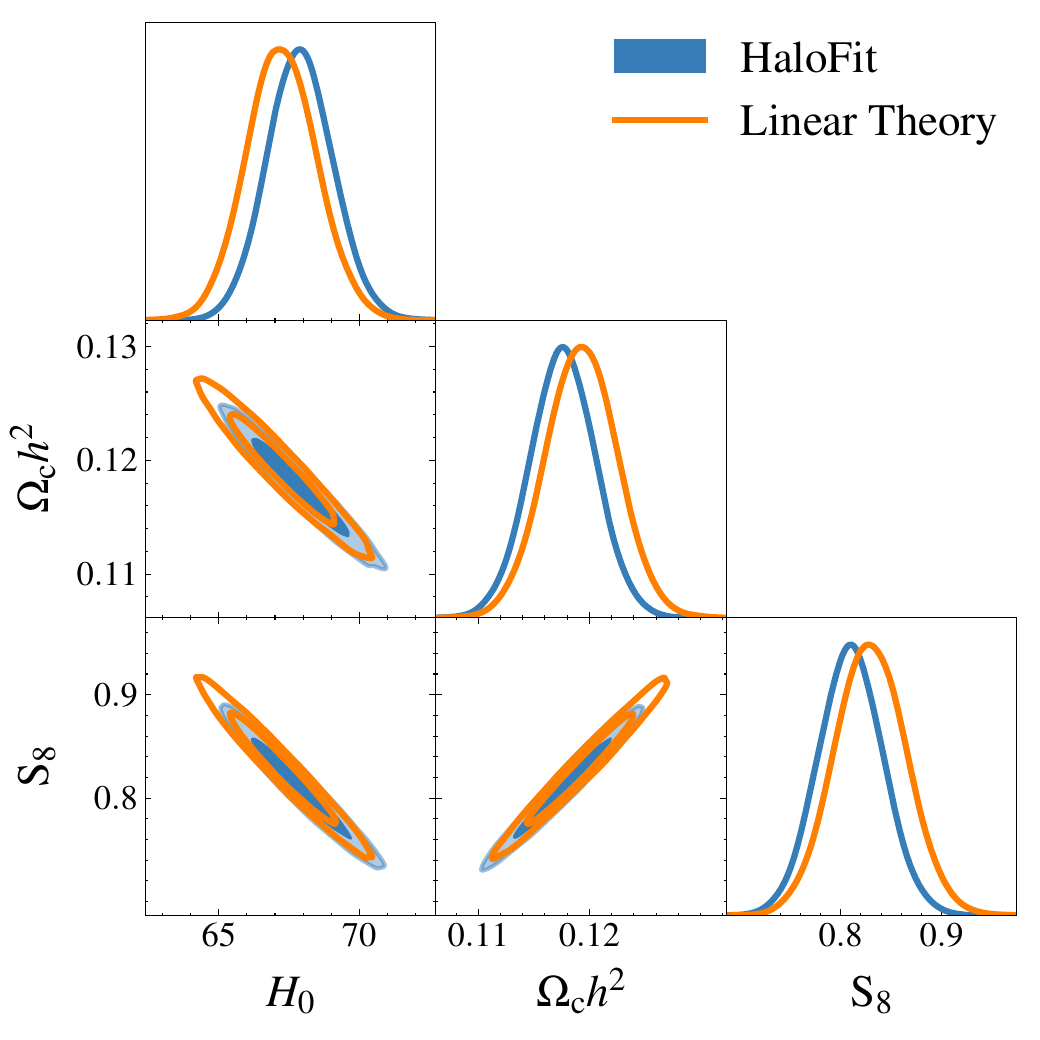} 
    \end{subfigure}
    \hfill
    \begin{subfigure}{0.45\textwidth}
        \includegraphics[width=\textwidth]{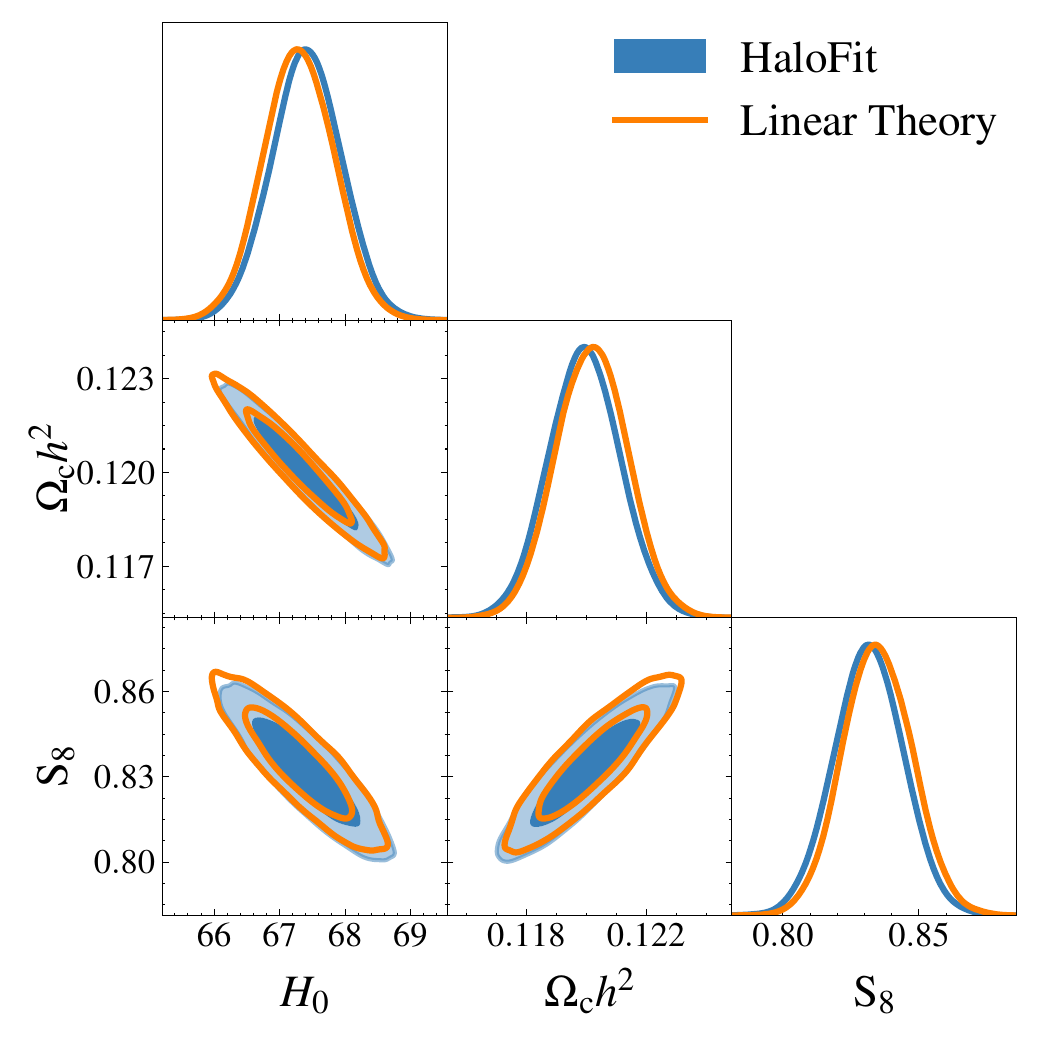}
    \end{subfigure}
    \caption{Marginalized 1D and 2D posteriors for $H_0,\ \Omega_ch^2$, and $S_8$ for $\Lambda$CDM when including non-linear corrections from \texttt{HaloFit} (blue) or when excluding them (orange). Left: using SPT + SPTlens data. Right: using Planck + Plancklens data.
    We observe a bias of $\mathcal{O}(0.6\sigma)$ in these three parameters when using SPT data and neglecting non-linear corrections.
    }
    \label{fig:LCDM_bias}
\end{figure}

However, when we include lensing data for {\it{Planck}} or SPT-3G, we notice larger shifts in parameters, particularly for $\Omega_ch^2,\ H_0$, and $S_8$ with SPT-3G data, as shown in Table~\ref{table:bias}. At face value, these quantities seem small, however it is crucial to emphasize that these biases are not simple statistical fluctuations. Since in both cases when non-linear corrections are included or excluded we are using the same data set, these shifts represent the bias that results from the absence of a correct theoretical description of non-linear corrections.
The marginalized 1D and 2D posteriors for these parameters are shown on the left hand side (l.h.s) of Figure~\ref{fig:LCDM_bias} for SPT+SPTlens, and on the right hand side (r.h.s) of the figure for Planck+Planklens. We note that each of these parameters shift in the same direction whether we test the {\it{Planck}} or SPT-3G datasets.
For the ACT DR6 lensing likelihood, we do not run our own chains; the authors in~\cite{ACT:2023kun} and~\cite{ACT:2023dou} show that linear theory shifts the central values of the $\sigma_8$ and $S_8$ parameters by $\sim0.6\sigma$, and we use this result as our point of reference for this dataset.

For $\Lambda$CDM, and a few of its extensions, the correct analysis would be to include the known non-linear corrections when constraining them. However, if non-linear corrections for more complicated models are not estimated, and hence are not included when constraining them, a bias of $\mathcal{O}(0.6\sigma)$ on $\Omega_ch^2,\ H_0$, or $S_8$ could completely change conclusions. For example, if a model claims a reduction in the Hubble tension to the level of $2.6\sigma$ when using SPT-3G data, this would not be accurate if there was an $\mathcal{O}(0.6\sigma)$ upward shift resulting from not including the appropriate non-linear corrections to this model.

Nevertheless, lensing data from CMB is very helpful in constraining cosmological models. We show this for the special case of interacting DM models, after their brief introduction in the next section.

%%%%%%%%% Bias Table %%%%%%%%%%
\begin{table}[htbp!]
\renewcommand{\arraystretch}{1.2}
\begin{center}
 \begin{tabular}{l @{\hskip 12pt} |@{\hskip 12pt} c @{\hskip 12pt} |@{\hskip 12pt} c @{\hskip 12pt}  } 
 \toprule
   Parameter(s) & Planck + Plancklens ($D$)
   & SPT + SPTlens ($D$)
   \\ [0.5ex] 
 \hline
$\Omega_\mathrm{c} h^2$ &  $0.17$ & $0.57$ \\

$H_0$ &  $0.20$ & $0.55$ \\
$S_8$  &   $0.23$ & $0.60$ \\ 
$\{\Omega_ch^2,H_0,S_8\}$ & $0.26$ & $0.62$ \\

  \hline
\end{tabular}
    \caption{
    The three individual parameter biases, eq.~\ref{eq:bias1}, along with $D$ of eq.~\ref{eq:biasN} for the combination of the three parameters, when non-linear corrections are not included in $\Lambda$CDM. These are computed for data sets Planck + Plancklens and SPT + SPTlens, as defined in Section~\ref{Sec:Data_Sets}.
    }
\label{table:bias}
\end{center}
\end{table}

\section{Dark Sector Models}
\label{sec:Models}

We consider two types of dark sector interactions, as implemented in \texttt{CLASS}\cite{2011JCAP...07..034B} according to the formalism presented in~\cite{Becker:2020hzj}, where the authors use their implementation to place constraints on these models using {\it{Planck}} 2018 temperature, polarization, and lensing data. In all cases, there are two parameters which are always relevant, regardless of the specific model: the fraction of DM which participates in the interaction, $f_\mathrm{IDM}$, and its mass, $m_\mathrm{IDM}$.

\subsection{Dark Matter Interacting with Baryons}
\label{subsec:DMB}

The first class of interactions that we consider is the case of DM scattering with baryons via the process DM + B $\leftrightarrow$ DM + B. Strong bounds on these interactions exist from other experiments, such as ground-based direct detection searches~\cite{ParticleDataGroup:2024cfk}. CMB data offer a complementary probe, since we require no assumptions about the details of the scattering with the atomic nucleus being used in the experiment, nor do we require assumptions about our local DM distribution. Furthermore, CMB searches can extend our sensitivity to lower mass ranges than what is probed by direct detection. The implementation in \texttt{CLASS} used in this work is valid so long as the DM particles are non-relativistic, which is the case for $m_\mathrm{IDM} \geq 1\ \mathrm{MeV}$.

The DM can in principle scatter from either protons or electrons, however the scattering cross-section with electrons is suppressed by the smaller electron mass, resulting in weaker bounds on the scattering cross-section~\cite{Li:2022mdj}. In this work we study only the scattering of DM with hydrogen atoms.
The scattering cross-section can be parameterized as
\beq
\sigma = \sigma_{\mathrm{DM-b}} v^{n_\mathrm{b}},
\eeq
where the exponent of the relative velocity, $n_{\mathrm{b}}$, captures the temperature dependence of the interaction and in weakly-coupled theories can take on even values. As $n_{\mathrm{b}}$ takes on larger values, the interaction decouples earlier and becomes less relevant at lower temperatures, thus we consider values of $n_{\mathrm{b}} \in \{-4, -2, 0\}$ as in~\cite{Becker:2020hzj}. 
Note that positive values of $n_b$ are possible, but they have considerable impact at very high temperatures prior to recombination.

\begin{figure}[]
    \centering
    \includegraphics[width=1.00\columnwidth]{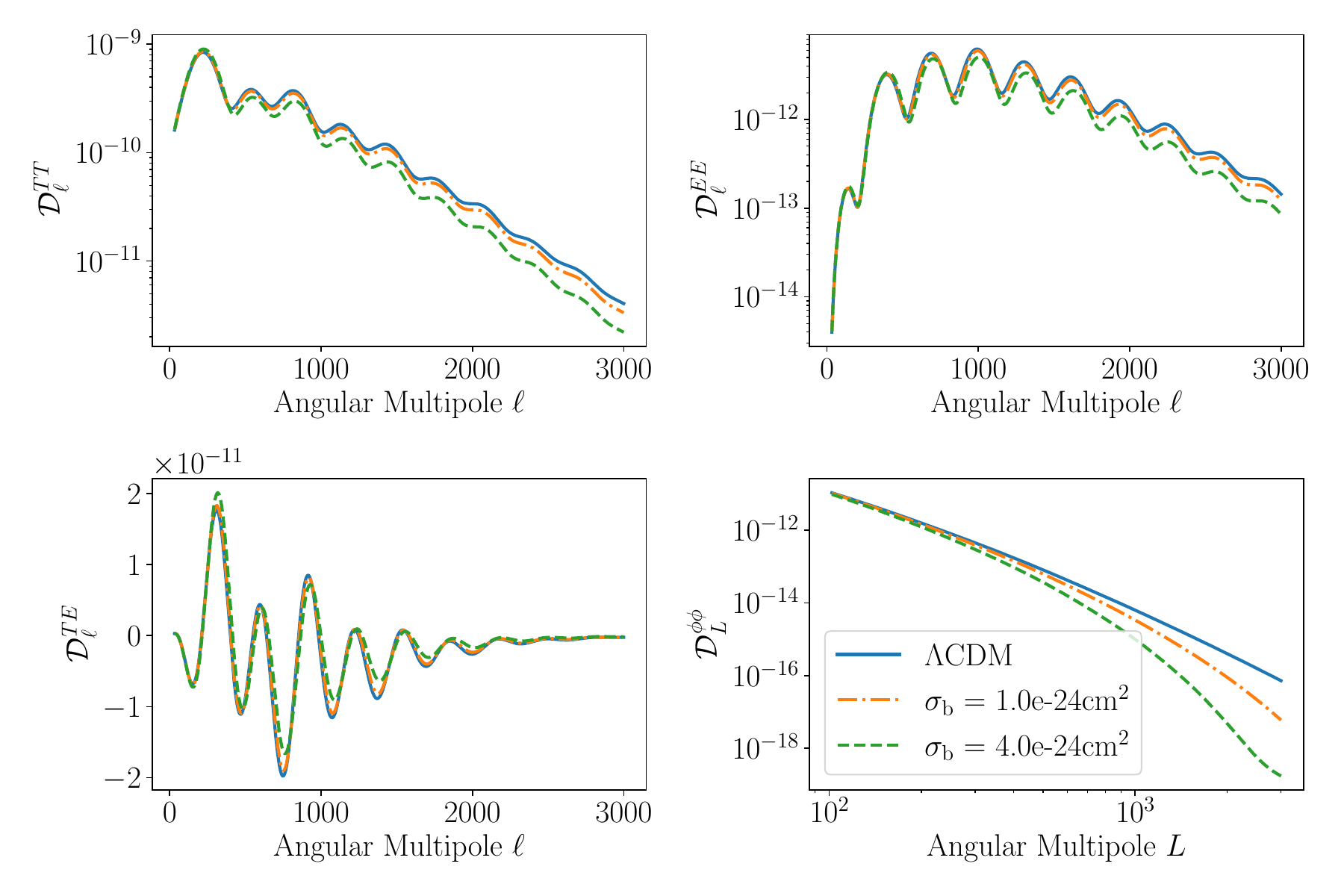}
    \caption{Dimensionless lensed CMB primary and lensing spectra, $\mathcal{D}_{\ell}$, for DM-b scattering with $n_{\mathrm{b}} = 0$ compared to $\Lambda$CDM (green). The blue and orange curves correspond to setting $\sigma_{\mathrm{b}}$ to $4 \times 10^{-24} \mathrm{cm}^2$ and $1 \times 10^{-24} \mathrm{cm}^2$, respectively. Note that these cross-sections are chosen to be an order of magnitude higher than current CMB bounds in order to facilitate visual clarity. 
    The scattering shifts the locations of the acoustic peaks in the primary CMB and suppresses small-scale power in the primary CMB and lensing power spectrum.
    }
    \label{fig:cmb_dmb}
\end{figure}

In Figure~\ref{fig:cmb_dmb}, we show lensed CMB temperature and polarization power spectra and lensing power spectra for a fiducial $\Lambda$CDM model compared to the case with DM-b interactions included. We plot the spectra
\be
\mathcal{D}_{\ell}^{X}=\frac{\ell(\ell+1)}{2\pi}C_{\ell}^{X}
\ee
where $C_{\ell}^{X}$ is the dimensionless CMB spectrum for $X\in\{TT,TE,EE,\phi\phi\}$.
The scattering cross-sections are set one order of magnitude higher than current CMB bounds to facilitate visual clarity of the impact on CMB power spectra. The result of the additional scattering interaction is a phase shift of the acoustic peaks of the primary CMB, and a suppression of power on small scales in both the primary anisotropies and the lensing power, both of which are important for constraining this model. 

%%%% DMB Figure %%%%%%%%%
%%%%%%%%%%%%%%%%

\subsection{Dark Matter Interacting with Dark Radiation}
\label{subsec:DMDR}

The second type of interaction we consider\footnote{Note that there is a third type of interaction, that of DR self-interactions, which was recently constrained in the context of Hubble Tension solutions~\cite{Khalife:2023qbu}. For brevity, we do not consider this kind of interaction here.} is an elastic scattering of DM with massless particles in the dark sector, often referred to as dark radiation (DR), via DM + DR $\leftrightarrow$ DM + DR. This type of interaction was proposed as a solution to the missing missile problem~\cite{Missing_Missiles2}, and described with different particle physics scenarios (see~\cite{Archidiacono:2019wdp,Cyr-Racine:2013fsa} and references therein for more details). However, as pointed out in~\cite{Archidiacono:2017slj}, within the general ETHOS formalism~\cite{Cyr-Racine:2015ihg} currently implemented in \texttt{CLASS}, the detailed particle physics mechanism has no direct impact on large-scale structure formation. 

In the ETHOS formalism, there are four parameters in the DM-DR interaction that are most relevant, in addition to $f_\mathrm{IDM}$ and $m_\mathrm{IDM}$\footnote{Note that, in principle, these two could be set as free parameters. However, in Section~\ref{Sec:Results}, we do not consider this scenario to avoid additional complexities.}. The first is the amount of DR relative to CMB photons, expressed as $\xi = T_\mathrm{DR}/T_{\gamma}$, where $T_i$ is the temperature of species $i=\{\text{DR},\gamma\}$. The second is the nature of DR, i.e. if it is free-streaming or self-interacting. The last two parameters are the amplitude of the interaction rate, $a_\mathrm{dark}$, and its redshift (i.e. temperature) dependence, $n_\mathrm{DR}$, respectively. These two appear in the expression for the interaction rate (see eq.(3.7) of~\cite{Archidiacono:2019wdp}):
\beq
\Gamma_\mathrm{DM-DR} = -\frac{4}{3}\omega_\mathrm{DR} a_\mathrm{dark} \left(\frac{1+z}{1+z_d}\right)^{1+n_\mathrm{DR}},
\label{Eq:DMDR_Scattering}
\eeq
where $\omega_{DR}\propto \xi^4$ is the physical density of the DR and $z_d$ is a normalization factor set to $10^7$, close to the expected decoupling redshift between DM and DR. Note that $n_\mathrm{DR}$ can take any value, with positive ones corresponding to a higher momentum exchange in the early Universe, while negative values produce more efficient momentum exchange in the late universe. 

In order to constrain DM-DR interactions, there are two parametrizations that have been widely used for a specific choice of DR nature and $n_\mathrm{DR}$. The first is to sample over $\xi$ and $\log_{10}(a_\mathrm{dark})$ (denoted $\log a_\mathrm{dark}$ below), which we label \texttt{param1}, while the second is on $a_\mathrm{dark}\xi^4$ and $\Delta N_\mathrm{eff} = N_\mathrm{eff}-3.044\propto\xi^4$, which we label \texttt{param2}, where $N_\mathrm{eff}$ is the effective number of relativistic degrees of freedom. In both types of parametrizations, previous works have placed bounds on these interactions for $n_\mathrm{DR} \in \{2, 4\}$~\cite{Archidiacono:2017slj,Archidiacono:2019wdp} and for $n_\mathrm{DR} =0$~\cite{Lesgourgues:2015wza,Buen-Abad:2017gxg,Archidiacono:2019wdp,Becker:2020hzj}. In the former two cases, DR self-interactions are subdominant~\cite{Archidiacono:2017slj}, while for the $n_\mathrm{DR}=0$ case they are comparable to the DM-DR strength. Hence, for $n_\mathrm{DR} \in \{2, 4\}$, DR will be free-streaming, while for $n_\mathrm{DR} =0$ it is self-interacting. Moreover, the latter case is of particular interest, as it maps onto a model which has been proposed to address the $H_0$ and $\sigma_8$ tensions~\cite{Lesgourgues:2015wza,Buen-Abad:2017gxg,Archidiacono:2019wdp,Becker:2020hzj}. 

\begin{figure}
		\begin{subfigure}[t]{0.5\textwidth}
			\centering
			\includegraphics[width=\textwidth]{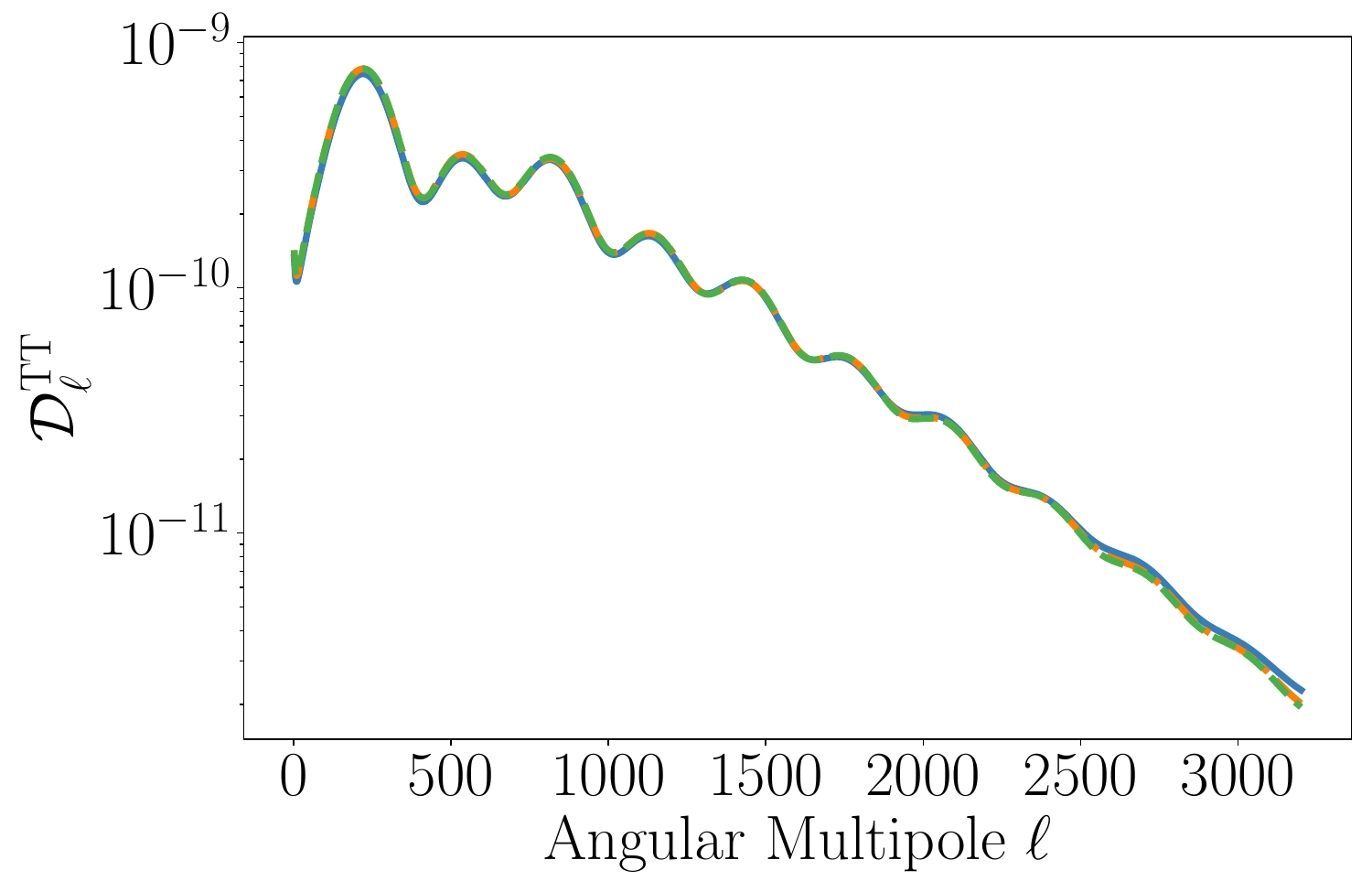}
		\end{subfigure}%
		\hfill
            \begin{subfigure}[t]{0.5\textwidth}
			\centering
			\includegraphics[width=\textwidth]{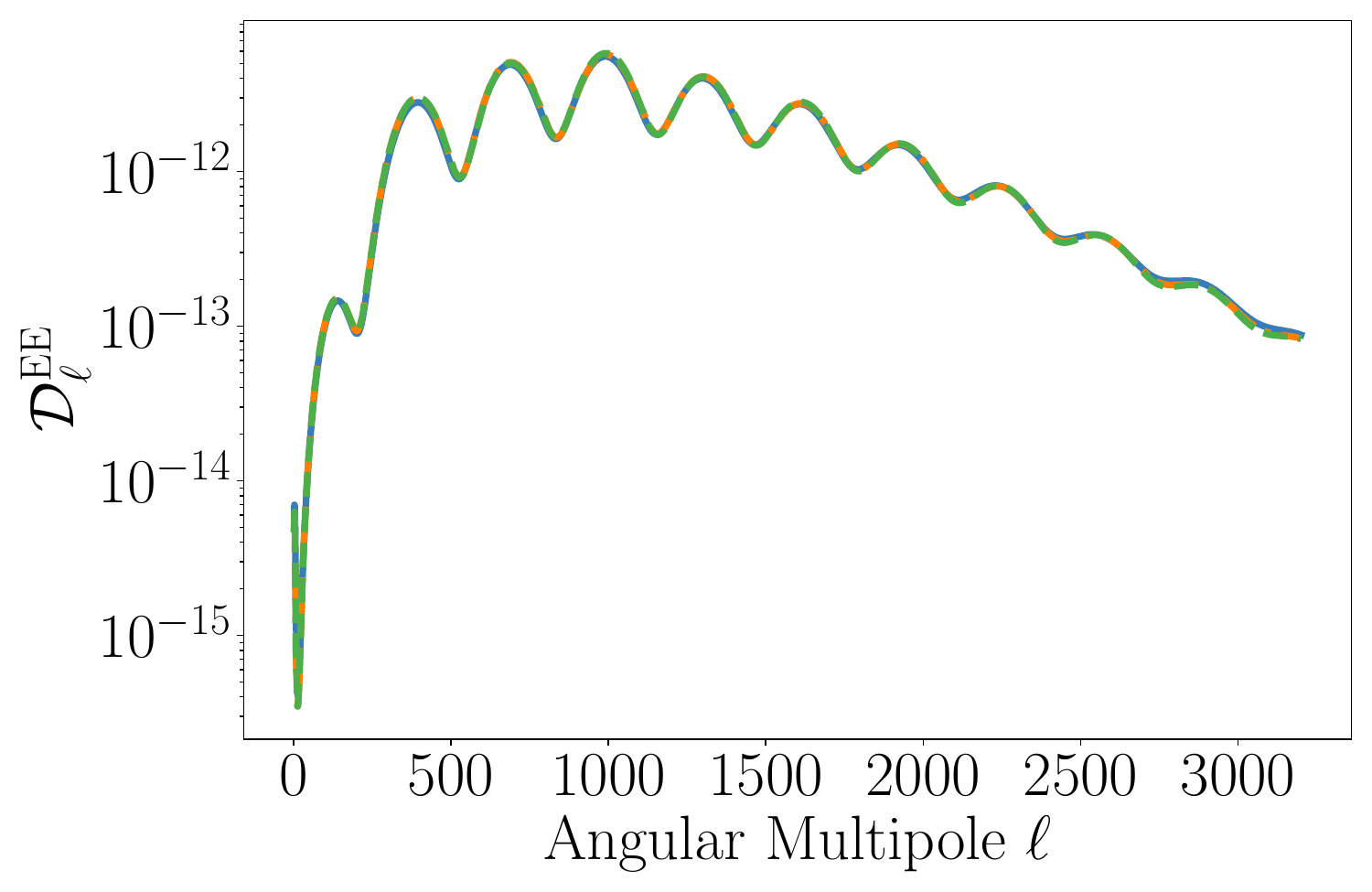}
		\end{subfigure}%
		\vspace{0.5cm}
		\begin{subfigure}[t]{0.5\textwidth}
			\centering
			\includegraphics[width=\textwidth]{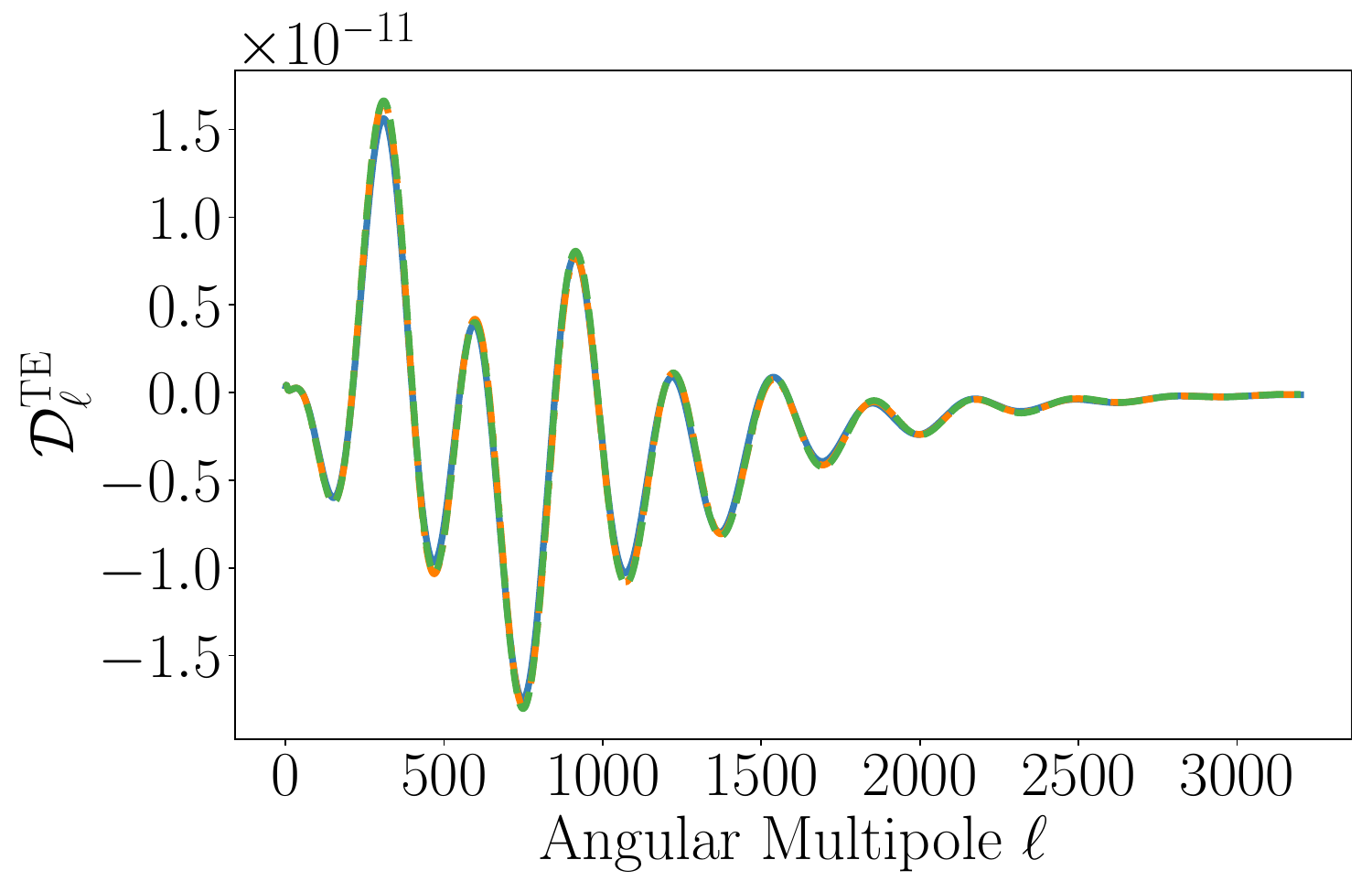}
		\end{subfigure}%
		\hfill
		\begin{subfigure}[t]{0.5\textwidth}
			\centering
			\includegraphics[width=\textwidth]{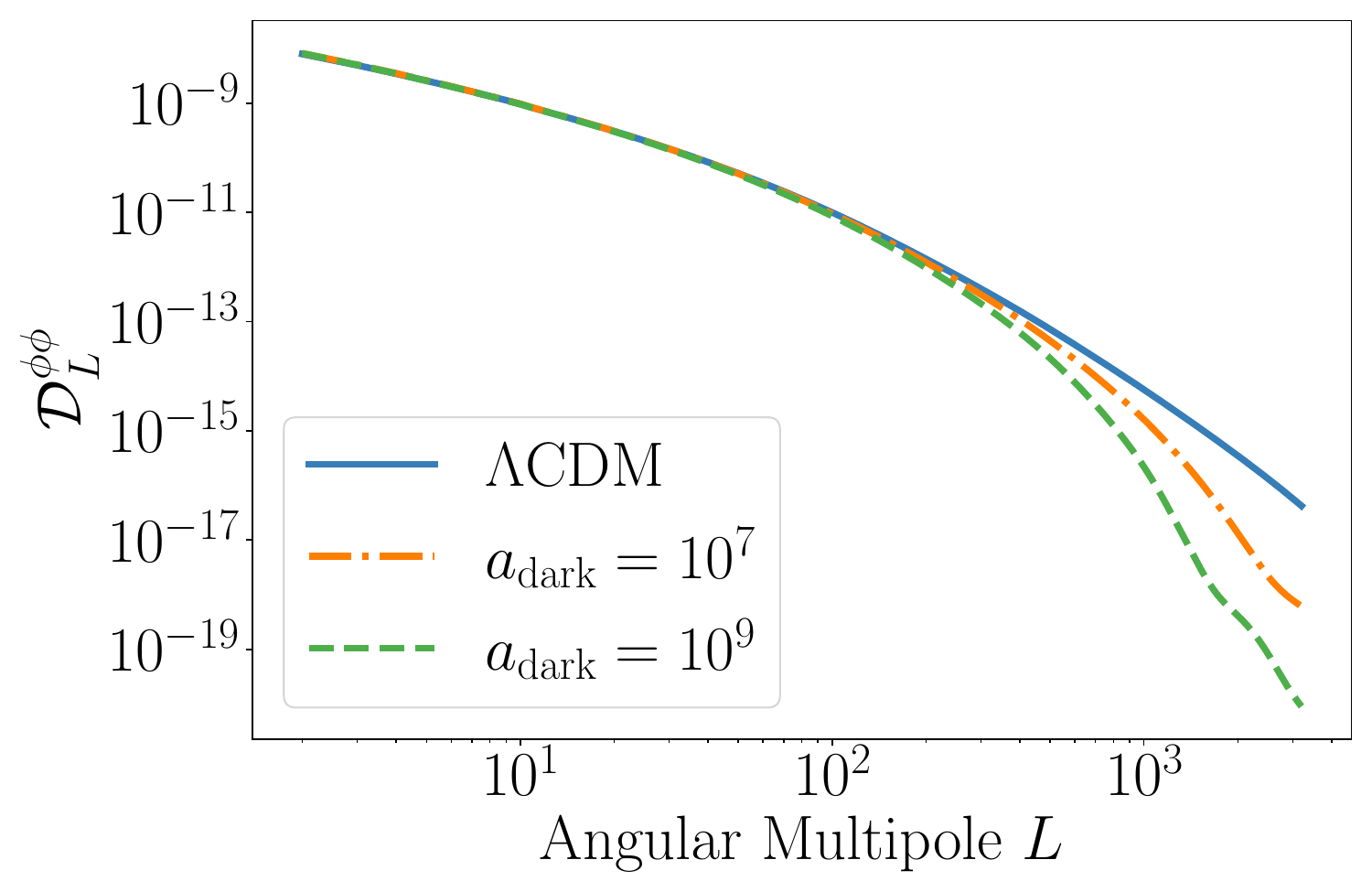}
		\end{subfigure}
		\caption{Dimensionless lensed CMB primary and lensing spectra, ${\cal D}_{\ell}$, for DM-DR with $n_\mathrm{DM-DR}=2$ compared to $\Lambda$CDM (green). The blue and orange curves correspond to setting $a_\mathrm{dark}$ in eq.~\eqref{Eq:DMDR_Scattering} to $10^7$ and $10^9$, respectively. The six $\Lambda$CDM parameters and $\xi$ are set consistently with the results of~\cite{Archidiacono:2019wdp}.
        The largest impact of this interaction is a suppression of lensing power at small scales, where nonlinear corrections to the matter power spectrum become important.
        }
\label{fig:cmb_dmdr}
	\end{figure}

As an illustration, we show in Figure~\ref{fig:cmb_dmdr} the lensed CMB temperature and polarization power spectra and lensing power spectra for a fiducial $\Lambda$CDM model compared to the case when DM-DR interactions with $n_\mathrm{DM-DR}=2$ are included. We choose two representative values of $a_\mathrm{dark}$, one near current CMB bounds ($a_\mathrm{dark}=10^7$), and one which is much larger ($a_\mathrm{dark}=10^9$), in order to better emphasize the changes to power spectra. This type of interaction has a minor effect on the primary CMB spectra, with only a slight shift in peak locations and a minor suppression of power at small scales ($\ell>2500)$ in $TT$ (top left plot). However, the major impact this interaction has is on the CMB lensing power spectrum, as is evident from the bottom right plot of Figure~\ref{fig:cmb_dmdr}. This hints at the importance of CMB lensing in constraining this type of interaction, which we show explicitly in Section~\ref{Sec:Results}.

%%%% DMDR Figure %%%%%%%%%
%%%%%%%%%%%%%%%%

\section{Matter Power Spectrum with Dark Sector Models}
\label{sec:Pk}

\begin{figure}[]
    \centering
    \includegraphics[width=1.00\columnwidth]{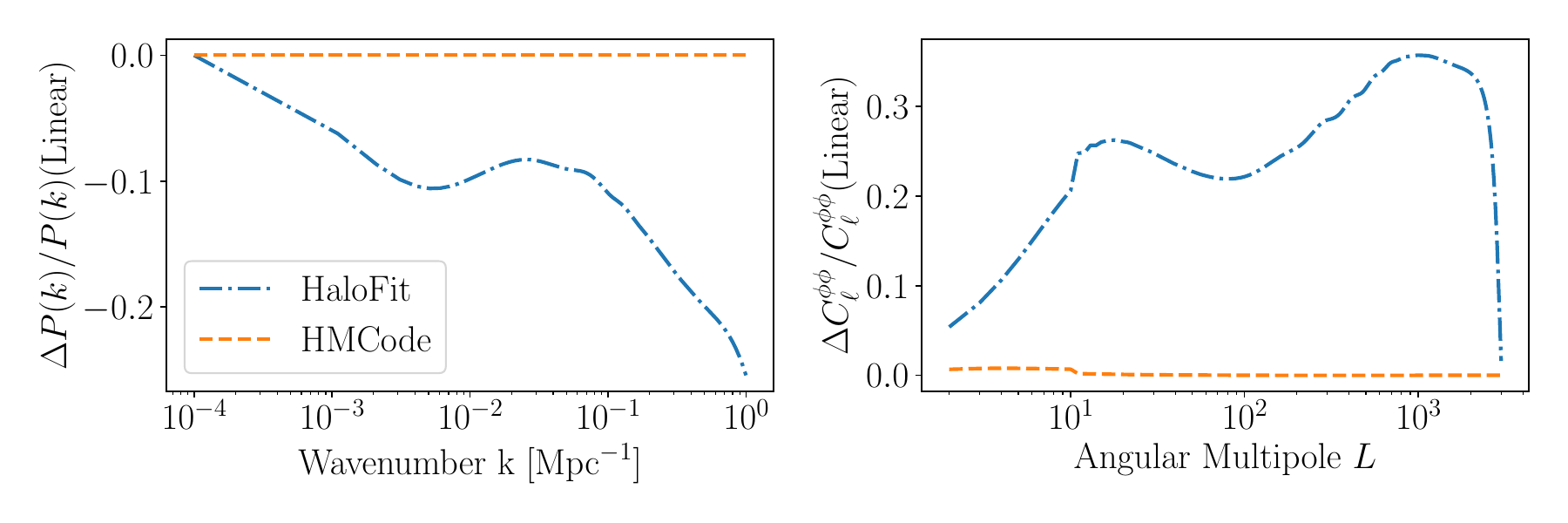}
    \caption{
    The difference in \texttt{CLASS} output between using linear theory and including non-linear corrections from \texttt{HaloFit} (blue) or \texttt{HMCode} (orange), for DM-b interactions with $n_\mathrm{b}=0$. 
    The model parameters used to generate the plot are: $m_\mathrm{IDM} = 1\ \mathrm{GeV}$, $f_\mathrm{IDM} = 1$, and $\sigma_\mathrm{DM-b} = 2.5 \times 10^{-25}\mathrm{cm}^{2}$.
    Left: fractional difference in the matter power spectrum $P(k)$ at redshift $z=1$.
    Right: fractional difference in the CMB lensing power spectrum $C_{\ell}^{\phi\phi}$. These codes show numerical instability when DM-b interactions are turned on.
    }
    \label{fig:non-linear_Pk_dmb}
\end{figure}

\begin{figure}[]
    \centering
    \includegraphics[width=1.00\columnwidth]{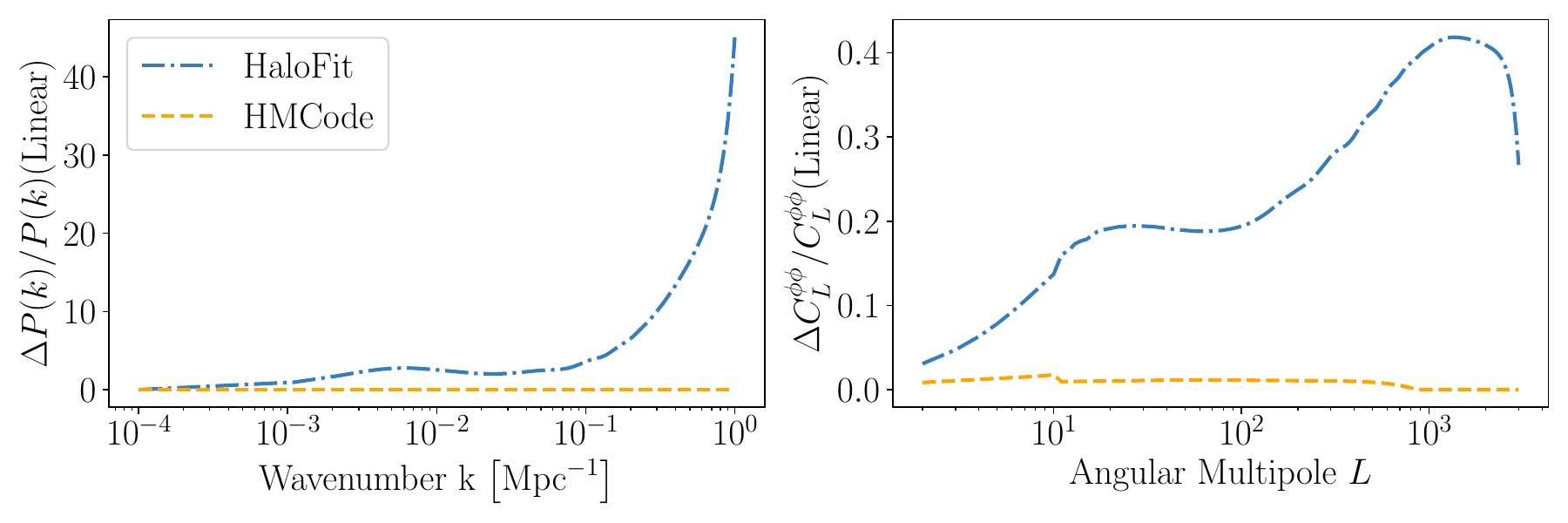}
    \caption{
    Same as Figure~\ref{fig:non-linear_Pk_dmb} for DM-DR interactions with $n_\mathrm{DM-DR}=2$. The model parameters used to generate the plot are: $m_\mathrm{IDM}=1$ GeV, $f_\mathrm{IDM}=1$, $\xi=0.17$, and $a_\mathrm{dark}=10^8$.
    }
    \label{fig:non-linear_Pk_dmdr}
\end{figure}

Figures~\ref{fig:cmb_dmb} and~\ref{fig:cmb_dmdr} have been computed without non-linear corrections to the matter power spectrum.
Indeed, \texttt{HaloFit} and \texttt{HMCode} have not been calibrated with the inclusion of extended dark sector models, such as those discussed in Section~\ref{sec:Models}.
In the l.h.s panels of Figures~\ref{fig:non-linear_Pk_dmb} and~\ref{fig:non-linear_Pk_dmdr}, we show the fractional difference in the matter power spectrum, $P(k)$, at a redshift of $z=1$ between linear theory and when non-linear corrections with \texttt{HMCode} (orange line) and \texttt{HaloFit} (blue line) are enabled at the same time as DM-b (Figure~\ref{fig:non-linear_Pk_dmb}) and DM-DR (Figure~\ref{fig:non-linear_Pk_dmdr}) interactions. Although one might hope to at least correctly approximate the non-linear physics of $\Lambda$CDM, the outputs are in fact unstable once additional interactions modify the matter power. This behavior translates into the CMB lensing power spectrum as well, shown in the r.h.s panels of Figures~\ref{fig:non-linear_Pk_dmb} and~\ref{fig:non-linear_Pk_dmdr}.
We note that we have chosen example interaction strengths which are near current CMB upper limits, making this behavior indeed relevant for current analyses. 

%%%% Nonlinear P(k) DMB Figure %%%%%%%%%
%%%% Nonlinear P(k) DMDR Figure %%%%%%%%%

%%%%%%%%%%%%%%%%

Due to these constraints, if one wants to use CMB data to search for these classes of DM interactions, one can reliably utilize only lensed CMB and lensing power spectra predictions which have been computed using the linear matter power spectrum. 
As discussed in section~\ref{Sec:Data_Sets}, we limit ourselves to cosmological datasets which are not sensitive to the impact of $\Lambda$CDM physics on non-linear scales. If one were to analyze such datasets using theory code outputs which do not include the non-linear matter power spectrum corrections, this would result in biases in cosmological parameter inferences, with the size of these biases depending on the sensitivity of a given dataset. 
Moreover, the behavior of both \texttt{HaloFit} and \texttt{HMCode} shown in Figures~\ref{fig:non-linear_Pk_dmb} and~\ref{fig:non-linear_Pk_dmdr} points to their numerical instabilities when including DM interactions.
Nevertheless, even if we could reliably (i.e. without risks of numerical instabilities) compute non-linear $P(k)$ for $\Lambda$CDM physics when turning on dark sector interactions in our cosmological theory code, we are still lacking an implementation of these corrections based on N-body simulations which also account for extended models. In other words, we are at risk of biasing our inferences for extended model parameters if we cannot quantify the effects of these models on non-linear scales. We additionally note that due to degeneracies which are present between some $\Lambda$CDM parameters and some parameters of extended models, any bias in one can lead to a bias in the other.

As an example of the consequences of biases to extension parameters if their impact on non-linear scales is not accounted for, consider Figure~\ref{fig:cmb_dmb}; we see that DM-b scattering tends to suppress structure growth and thus CMB lensing power at small scales, while in general, non-linear growth tends to increase power. If one performs parameter fitting without accounting for the non-linear effects of the DM-b scattering, which could in principle be non-negligible, the likelihood will prefer cosmological parameter values which increase the small-scale power. This is equivalent to less DM-b scattering and thus the sampling would inherently have a preference for smaller cross-sections; this could result in a missed detection even if the data in principle have sensitivity down to the value of the cross-section.

\section{Results}
\label{Sec:Results}

In our analysis, we make the conservative choice to include only the datasets Planck, SPT, ACT, and BAO, with each dataset defined in Section~\ref{Sec:Data_Sets}. All model predictions from the theory code are computed with linear $P(k)$ theory only. As shown in Section~\ref{Sec:Nonlinear}, CMB lensing is more sensitive to the details of non-linear physics than primary CMB data, specially in the case of SPTlens. Quantifying the scales at which non-linearity from interactions within the dark sector becomes significant, along with the size of these effects, is beyond the scope of our current work. We also note that the SPT-3G 2018 TT/TE/EE data is new for the two models studied in this paper, while the ACT DR4 TT/TE/EE data is new for the DM-DR cases considered here. 
For all chains, we vary the six $\Lambda$CDM parameters~\eqref{eq:LCDM_params}, substituting $H_0$ with $100\theta_\mathrm{s}$, where $\theta_\mathrm{s}$ is the angular size of the sound horizon at recombination. We also model the neutrino sector as three massive species, each with mass $m_\nu = 0.02$ GeV.

\subsection{Dark Matter Interacting with Baryons}
\label{subsec:DMB-Results}

In~\cite{Becker:2020hzj}, the authors showed that, in the limit that $m_\mathrm{IDM} \gg m_\mathrm{b}$, a bound on $\sigma_\mathrm{DM-b}$ can be approximately rescaled as $\sigma_\mathrm{DM-b}(m_\mathrm{ref}/m_\mathrm{IDM})$ for some reference mass $m_\mathrm{ref}$, and allowing the mass to vary freely does not provide any meaningful constraint on its value. We therefore fix $m_\mathrm{IDM}=1$ GeV and allow $\sigma_\mathrm{DM-b}$ to vary for the three choices of $n_{\mathrm{b}} \in \{-4, -2, 0\}$.

%%%%%%%%% DM-Baryon Table %%%%%%%%%%
\begin{table}[htbp!]
\renewcommand{\arraystretch}{1.2}
\begin{center}
 \begin{tabular}{l @{\hskip 12pt} |@{\hskip 12pt} c @{\hskip 12pt} |@{\hskip 12pt} c @{\hskip 12pt} |@{\hskip 12pt}c@{\hskip 12pt}  } 
 \toprule
   Model & PB & PBS & PBSA   \\ [0.5ex] 
 \hline
$n_b = -4$ &  $2.9 \times 10^{-41}$ & $2.3 \times 10^{-41}$ 
& $3.1 \times 10^{-41}$  \\

$n_b = -2$ &  $3.8 \times 10^{-33}$ & $3.2 \times 10^{-33}$
& $3.8 \times 10^{-33}$	\\

$n_b = 0$  &   $2.4 \times 10^{-25}$ & $2.2 \times 10^{-25}$
& $2.1 \times 10^{-25}$  \\ 

  \hline
\end{tabular}
    \caption{
    95\% upper limits on DM-b scattering cross-section, $\sigma_\mathrm{DM-b}$, in units of $\mathrm{cm}^2$, for 
    $m_\mathrm{IDM} = 1\mathrm{}$ GeV, $f_\mathrm{IDM} = 1$,
    excluding CMB lensing data. Data set combinations are defined in Section~\ref{Sec:Data_Sets}.
    }
\label{table:dmb2sf}
\end{center}
\end{table}

In Table~\ref{table:dmb2sf}, we present 95\% CL upper bounds on the DM-b scattering cross-section for each fixed value of $n_\mathrm{b}$. The constraints from PB are consistent with results from~\cite{Becker:2020hzj}. The addition of the SPT-3G 2018 temperature and polarization data provides an improvement of order 10-20\%, depending on the velocity-dependence of the model. Although this is a concrete improvement on these upper limits, they are not as strong as what could be gained with the addition of lensing data. In all three cases considered here, we find that the scattering cross-section as constrained by the combination of Planck, SPT, and BAO data is consistent with zero.

\begin{figure}[]
    \centering
    \includegraphics[width=0.50\columnwidth]{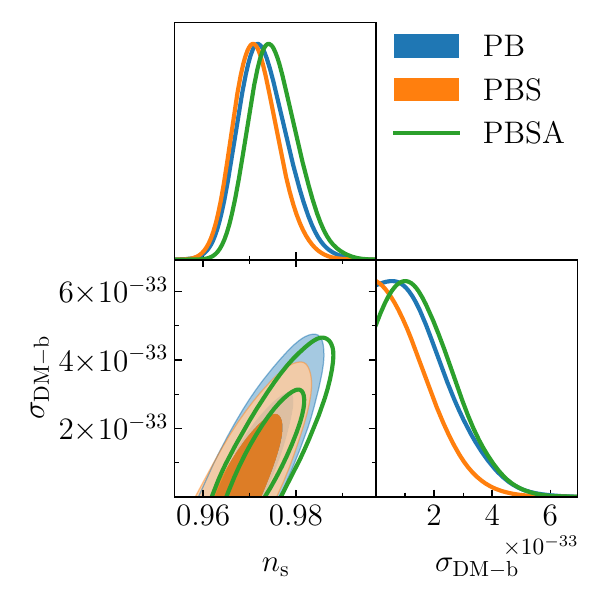}
    \caption{The 1D and 2D marginalized posteriors for $n_s$ and $\sigma_{\mathrm{DM-b}}$. These two parameters are noticeably degenerate; any shifts in the central values of $n_s$ can complicate the interpretation of upper bounds on $\sigma_{\mathrm{DM-b}}$.}
    \label{fig:dm_baryon_corner}
\end{figure}

With the further addition of the ACT DR4 temperature and polarization data, some of the upper bounds appear to weaken. In Figure~\ref{fig:dm_baryon_corner}, we show the 2D constraints on the DM-b scattering cross-section and $n_s$, a parameter which shows a strong degeneracy with $\sigma_\mathrm{DM-b}$. We note that the addition of the ACT data causes the one-dimensional marginalized posterior for $\sigma_\mathrm{DM-b}$ to shift slightly away from zero, thus increasing the 95\% upper limit even though the overall constraining power of the data combination is not weakening; similar behavior was observed in the results of~\cite{Li:2022mdj}. This example illustrates the point that any systematic biases that could shift the best-fit values of $\Lambda$CDM parameters could in turn complicate the interpretation of upper bounds on extended models, in situations where there are degeneracies between combinations of these parameters.

%%%% DM-Baryon Corner Plot %%%%%%%%%
%%%%%%%%%%%%%%%%

We do not perform any MCMC analyses using CMB lensing data to constrain DM-b scattering. However, in order to illustrate the constraining power contained in this lensing information, we compute Fisher forecasts for the DM-b scattering cross-section for the $n_\mathrm{b}=0$ model, from some recent and upcoming SPT-3G datasets. 
Fisher forecasting is performed with \texttt{FisherLens}~\cite{Hotinli:2021umk}, which includes the non-Gaussian covariance matrix to account for lensing-induced correlations between different multipoles, and using the lensed CMB power spectra.
The elements of the Fisher matrix are given by
\begin{equation}
    F_{ij} = \sum\limits_{\ell_1, \ell_2} \ \sum\limits_{W X Y Z} 
    \frac{\partial C_{\ell_1}^{XY}}{\partial \lambda^i} 
    \left[ \mathrm{Cov}_{\ell_1\ell_2}^{XY,WZ} \right]^{-1}
    \frac{\partial C_{\ell_2}^{WZ}}{\partial \lambda^j} \, 
    \label{eq:FisherMatrix}
\end{equation}
where $\lambda_{i}$ are cosmological parameters and $\mathrm{Cov}_{\ell_1\ell_2}^{XY,WZ}$ is the covariance matrix.
We include the multipole ranges $100 < \ell < 3000$ for $TT$, $100 < \ell < 3500$ for $TE$ and $EE$, and additionally $100 < L < 3000$ whenever lensing information is included, with lensing reconstruction utilizing $TT$ information up to $\ell$ of 3000.
We consider two data sets taken by SPT-3G: the first from observations taken in 2018, while the second from observations taken during the 2019-2020 observing seasons.
The foregrounds are modeled as in~\cite{SPT-3G:2024qkd}, assuming an internal frequency combination from different bands~\cite{Raghunathan:2023yfe}, with noise levels as in~\cite{SPT-3G:2021eoc} for the 2018 data and as in~\cite{SPT-3G:2024qkd} for the 2019-2020 data.
We assume a survey sky fraction of 3.6\% corresponding to the SPT-3G main winter field.
The cosmological parameter values and step sizes used for the computations are given in Table~\ref{table:cosmo_fiducial}. We use a $\tau$ prior of $\sigma_\tau = 0.007$. Neutrino masses are taken to have a sum of 60 meV, and the primordial helium abundance is set to be consistent with Big Bang nucleosynthesis predictions.

%%%%%%%%% Parameter Table %%%%%%%%%%
\begin{table}
\begin{center}
 \begin{tabular}{l @{\hskip 12pt}c@{\hskip 12pt}c} 
 \toprule
   %$\Lambda$CDM 
   Parameter    &   Fiducial Value      & Step Size     \\ [0.5ex] 
 \hline
$\Omega_\mathrm{c} h^2$ &   0.1197 	            & 0.0030 	    \\ 
$\Omega_\mathrm{b} h^2$ &   0.0222 	            & $8.0\times10^{-4}$ 	    \\
$\theta_\mathrm{s}$     &   0.010409 	            & $5.0\times10^{-5}$ 	    \\
$\tau$         &   0.060 	            & 0.020 	    \\
$A_\mathrm{s}$          &   $2.196\times10^{-9}$  & $0.1\times10^{-9}$ 	    \\
$n_\mathrm{s}$          &   0.9655 	            & 0.010 	    \\
$\sigma_\mathrm{DM-b}$          &   0 	            & $2 \times 10^{-26} (m_{\mathrm{IDM}}/\mathrm{GeV})^{0.15}$ 	    \\
  \hline
\end{tabular}
    \caption{
    The cosmological parameter values and step sizes (for numerical differentiation) used in Fisher matrix forecasts. The step sizes for $\Lambda$CDM parameters are taken from~\cite{Allison:2015qca}, while the step sizes for DM-b scattering are chosen according to~\cite{Li:2018zdm}.
    }
\label{table:cosmo_fiducial}
\end{center}
\end{table}
%%%%%%%%%%%%%%%%%%%%%%%%%%%%%%%

We present our forecasted parameter constraints in Table~\ref{table:dmbFisher}. Note that because the scattering cross-section is bounded below by zero, the posterior for this parameter is highly non-Gaussian and thus likely to be poorly approximated by the assumptions of Gaussianity inherent in the Fisher approximation. We therefore emphasize that these forecasts illustrate the relative value of different CMB datasets, but the absolute values of the forecasted error bars may differ from the results of numerically sampling the posterior.

We find that lensing information, which we have discarded for the SPT-3G 2018 dataset used in this analysis, can improve parameter constraints by about 30\%. Forecasts presented in~\cite{Li:2018zdm} showed that lensing information becomes the dominant source of constraining power once the sensitivity of a CMB experiment reaches 1$\mu$K. The improved noise levels of the SPT-3G 2020 observations demonstrate this point, as the relative value of the lensing information increases, offering a 50\% improvement in constraints on the cross-section. However, this dataset, as shown in~\cite{SPT-3G:2024atg}, is sensitive to the amplitude of non-linear structure growth at $> 3\sigma$. Without an accurate model of the non-linear physics, including the effects of DM-b interactions, using this lensing dataset would incur an unknown systematic error in the result, as discussed in Section~\ref{sec:Pk}. 
%%%%%%%%% DM-Baryon Fisher Forecasts %%%%%%%%%%
\begin{table}[htbp!]
\renewcommand{\arraystretch}{1.2}
\begin{center}
 \begin{tabular}{l @{\hskip 12pt} |@{\hskip 12pt} c @{\hskip 12pt}  } 
 \toprule
   Dataset used in forecast & 1-$\sigma$ error ($\mathrm{cm}^2$)  \\ [0.5ex] 
 \hline
SPT-3G 2018 TT/TE/EE &  $2.7 \times 10^{-25}$ \\

SPT-3G 2018 TT/TE/EE + Lensing &  $1.8 \times 10^{-25}$ \\

SPT-3G 2019-2020 TT/TE/EE   &   $1.9 \times 10^{-25}$ \\ 

SPT-3G 2019-2020 TT/TE/EE + Lensing   &   $9.5 \times 10^{-26}$ \\ 

  \hline
\end{tabular}
    \caption{
    Forecasted 1-$\sigma$ error for the DM-b scattering cross-section in the $n_\mathrm{b}=0$ model. Because the posteriors for this parameter are non-Gaussian, Fisher forecasts illustrate the relative constraining power of different datasets but may otherwise differ from the values that would be obtained by numerically sampling the posteriors. 
    The inclusion of lensing datasets would provide a 30-50\% improvement in constraining this parameter.
    }
\label{table:dmbFisher}
\end{center}
\end{table}

\subsection{Dark Matter Interacting with Dark Radiation}
\label{subsec:DMDR-Results}

In this section, we present our constraints on DM-DR interactions with $n_\mathrm{DR}\in\{2,4\}$, corresponding to free-streaming DR. We use \texttt{param1} and \texttt{param2} parametrizations introduced in Section~\ref{subsec:DMDR}. We do not consider the case $n_\mathrm{DR}=0$ as this case has been studied more exhaustively~\cite{Archidiacono:2019wdp,Becker:2020hzj,SPT_DES_DMDR}, even though it suffers (to a lesser extent) from the constraining difficulties we aim to highlight in this work. In all these runs, we set $f_\mathrm{IDM}=1$ and $m_\mathrm{IDM}=1$ GeV\footnote{As in the case for DM-b, as long as the $m_\mathrm{IDM}>1$ MeV, the exact value of the mass does not impact the dynamics.} and use uniform priors on the DM-DR parameters.

\begin{figure}[h!]
    \centering
    \begin{subfigure}{0.45\textwidth}
        \includegraphics[width=\textwidth]{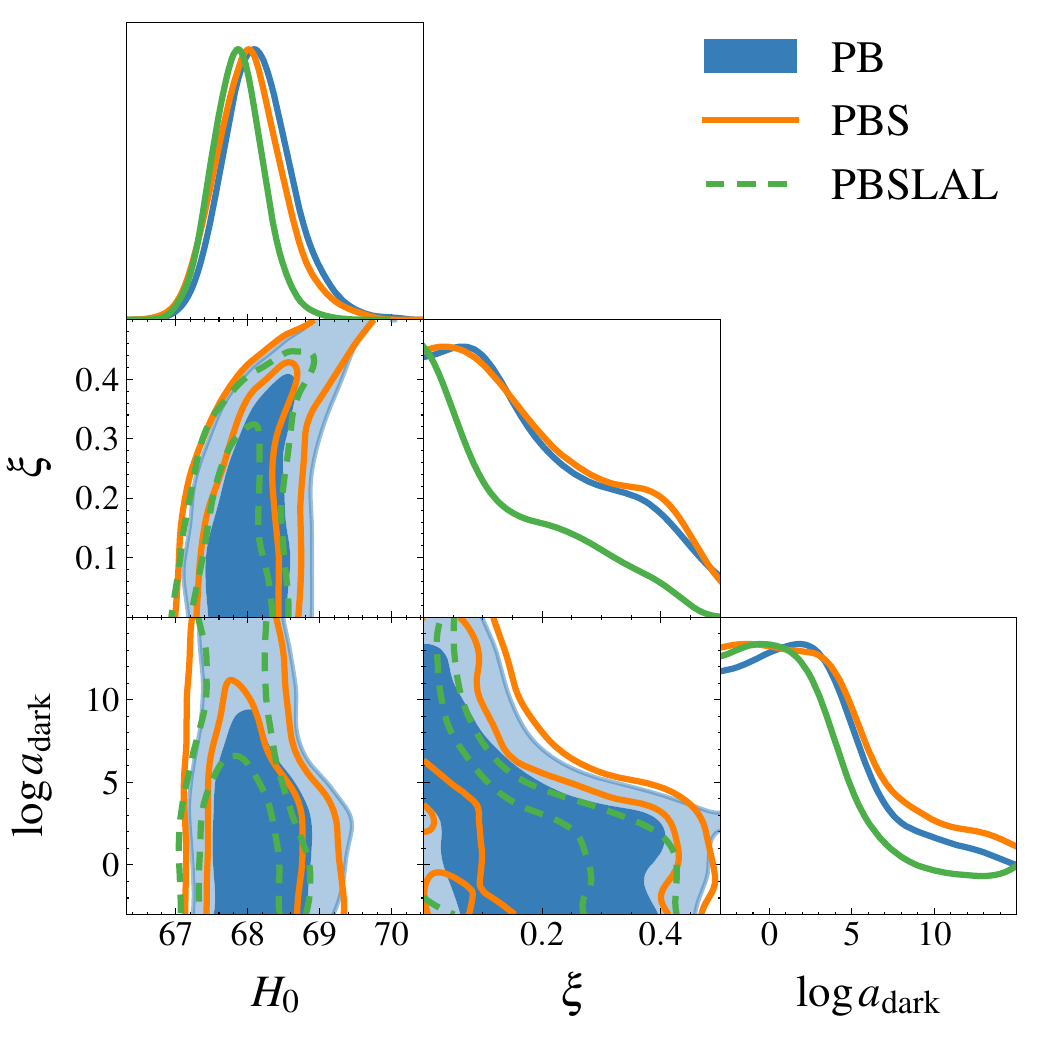} 
    \end{subfigure}
    \hfill
    \begin{subfigure}{0.45\textwidth}
        \includegraphics[width=\textwidth]{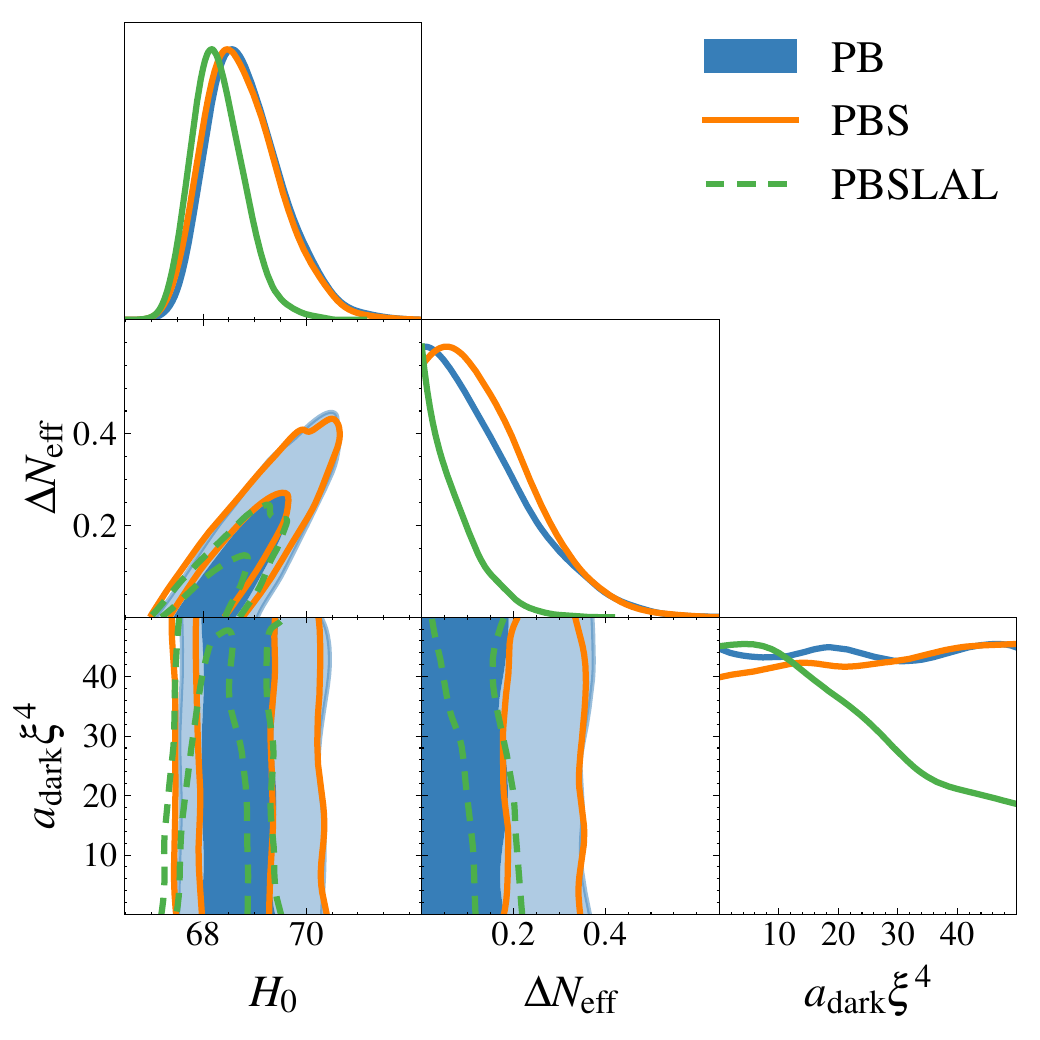}
    \end{subfigure}
    \caption{
    Marginalized 1D and 2D posteriors for $H_0$ and the DM-DR interaction parameters for the case $n_\mathrm{DM-DR}=2$.
    Left: using \texttt{param1}. Right: using \texttt{param2}. 
    The different data set combinations are defined in Section~\ref{Sec:Data_Sets}. 
    The constraints from the PBSLAL data combination are to be taken with caution, as they use data sensitive to non-linear corrections. Adding lensing information from $Planck$, SPT-3G, and ACT improves the constraints on DM-DR interactions when using \texttt{param1}, while it mildly does so when using \texttt{param2}.
    }
    \label{fig:DMDR_N2}
\end{figure}

\begin{figure}[h!]
    \centering
    \begin{subfigure}{0.45\textwidth}
        \includegraphics[width=\textwidth]{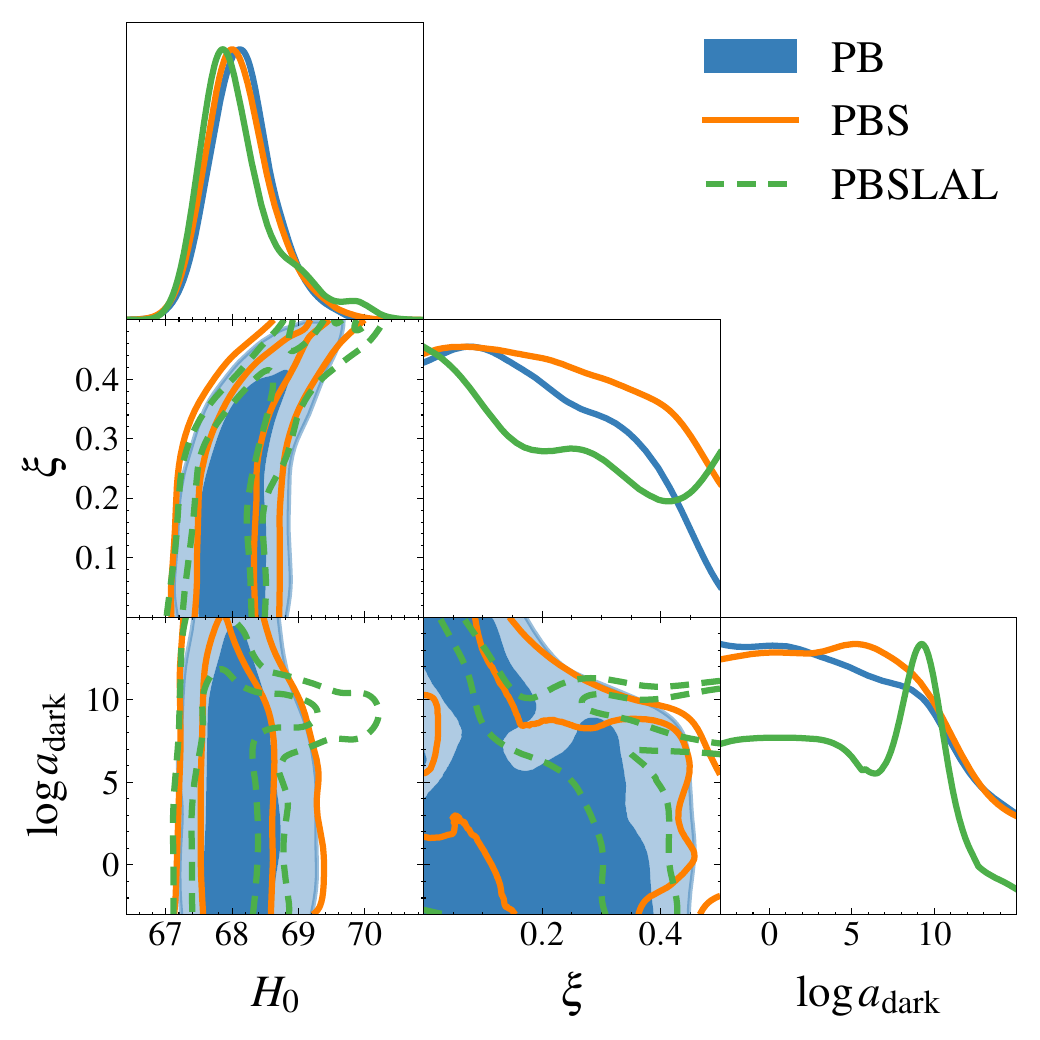} 
    \end{subfigure}
    \hfill
    \begin{subfigure}{0.45\textwidth}
        \includegraphics[width=\textwidth]{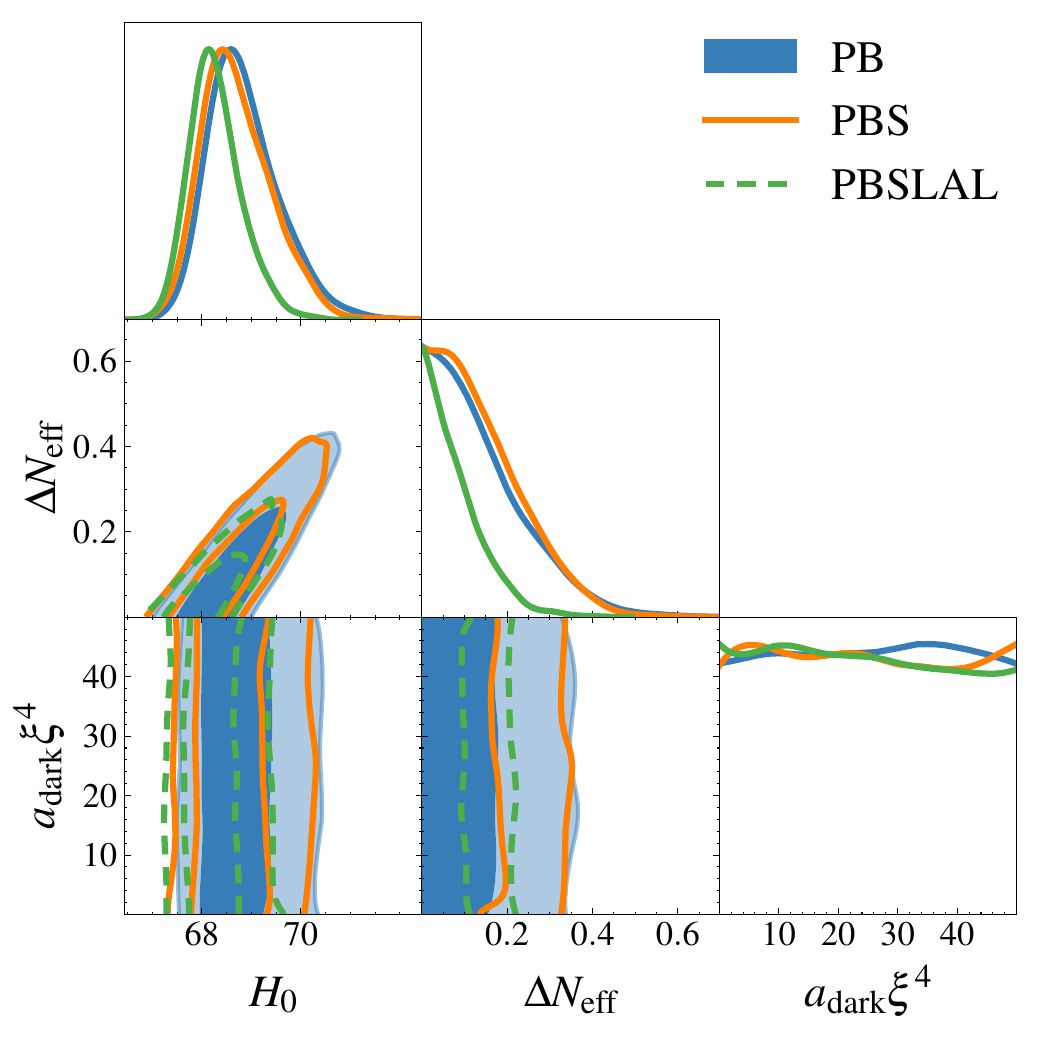}
    \end{subfigure}
    \caption{
    Same as Figure~\ref{fig:DMDR_N2} for $n_\mathrm{DM-DR}=4$. 
    The constraints from the PBSLAL data combination are to be taken with caution, as they use data sensitive to non-linear corrections.}
    \label{fig:DMDR_N4}
\end{figure}

\begin{table}[htbp!]
    \centering
    \begin{tabular}{l |c c c c |c c c c}
    \cline{2-9}
    & \multicolumn{4}{c|}{$n_\mathrm{DM-DR}=2$} & \multicolumn{4}{c}{$n_\mathrm{DM-DR}=4$} \\
        \hline
          Data Sets & $\xi$ & $\log_{10}(a_\mathrm{dark})$ & $a_\mathrm{dark}\xi^4$ & $\Delta N_\mathrm{eff}$ & $\xi$ & $\log_{10}(a_\mathrm{dark})$ & $a_\mathrm{dark}\xi^4$ & $\Delta N_\mathrm{eff}$\\
         \hline
         PB & n.l & n.l & n.l & $<0.36$ & n.l & n.l & n.l & $<0.35$\\
         PBS & n.l & n.l & n.l & $<0.35$ & n.l & n.l & n.l & $<0.34$\\
         PBSLAL$^*$ & $<0.39$ & $<8.2$ & n.l & $<0.19$ & $<0.40$ & $<12.30$ & n.l & $<0.21$\\
         \hline
         Priors & $\mathcal{U}(0,0.5)$ & $\mathcal{U}(-3,15)$ & $\mathcal{U}(0,50)$ & $\mathcal{U}(0,2)$ & $\mathcal{U}(0,0.5)$ & $\mathcal{U}(-3,15)$ & $\mathcal{U}(0,50)$ & $\mathcal{U}(0,2)$\\
    \end{tabular}
    \caption{Upper limits at 95\% CL on DM-DR interactions parameters \texttt{param1}$=\{\xi, \log_{10}(a_\mathrm{dark})\}$ and \texttt{param2}$=\{a_\mathrm{dark}\xi^4,\Delta N_\mathrm{eff}\}$ in the two cases $n_\mathrm{DM-DR}=2$ (columns 2-5) and $n_\mathrm{DM-DR}=4$ (columns 6-9). Uniform priors between $a$ and $b$, denoted as $\mathcal{U}(a,b)$, are used for every parameter. Most of these runs have no limit (n.l), and are thus unconstrained, unless lensing information in included. $^*$ Constraints are to be taken with caution, as they use data sensitive to non-linear corrections.}
    \label{tab:DMDR_Constraints}
\end{table}

Starting with parametrization \texttt{param1}, the l.h.s plots of Figures~\ref{fig:DMDR_N2} and~\ref{fig:DMDR_N4} show that primary CMB and BAO data (blue and orange curves) alone do not constrain well this model with either $n_\mathrm{DM-DR}=2$ or $n_\mathrm{DM-DR}=4$, respectively. Only when adding lensing data (green curve) we get $\xi < 0.39$ and $\log_{10}(a_\mathrm{dark}) < 8.2$ at 95\% CL for $n_\mathrm{DM-DR}=2$, and $\xi < 0.40$ and $\log_{10}(a_\mathrm{dark}) < 12.30$ at 95\% CL for $n_\mathrm{DM-DR}=4$ (Table~\ref{tab:DMDR_Constraints}). However, as mentioned before, these constraints cannot be trusted since the non-linear corrections for this model were not included. This shows again the importance of estimating non-linear corrections for these models in order to extract meaningful constraints.

On the other hand, with the parametrization \texttt{param2}, even including lensing data does not constrain this model with either $n_\mathrm{DM-DR}=2$ or $n_\mathrm{DM-DR}=4$. This can be seen from the right hand side (r.h.s) plots of Figures~\ref{fig:DMDR_N2} and~\ref{fig:DMDR_N4}, in addition to Table~\ref{tab:DMDR_Constraints}.\footnote{Note that the upper limit on $\Delta N_\mathrm{eff}$ comes from the CMB's sensitivity to additional relativistic degrees of freedom, which is not specific to the models considered here.} This parametrization is best constrained with information from Lyman-$\alpha$ forest flux, as presented in~\cite{Archidiacono:2019wdp}. However, this type of information cannot be used without running N-body simulations or using a parametrized description of the matter power spectrum~\cite{Lyman_parm1,Lyman_parm2,Lyman_parm3}. This again highlights the need to estimate non-linear corrections for such models if one is to extract meaningful constraints on them from CMB lensing or Lyman-$\alpha$ data.

\section{Conclusion}
\label{Sec:Conclusion}
In this article, we aim to caution the use of current and upcoming CMB lensing data in constraining beyond $\Lambda$CDM models without estimating their non-linear corrections. We show this in two ways.

First, in Figure~\ref{fig:LCDM_bias}, we show the bias in parameter constraints within $\Lambda$CDM when non-linear corrections are completely ignored. The $\sim \mathcal{O}(0.6\sigma)$ shift in parameters when using SPT-3G 2018 lensing data without including non-linear corrections is a representation of the amount of bias that results from completely ignoring these corrections. Given that the latter strategy is often used when constraining models beyond $\Lambda$CDM, the reported bias in this work is to be taken into account when interpreting those constraints.

Second, we consider two interacting DM models as a case study for the importance of CMB lensing data in constraining models. The first is DM interacting with baryons (DM-b), while the second is DM interacting with DR (DM-DR). For each type of interaction, we show the instability of currently available estimation methods of non-linear corrections in computing the matter and lensing power spectra (Figures~\ref{fig:non-linear_Pk_dmb}-~\ref{fig:non-linear_Pk_dmdr}). This performance justifies not using these corrections when interacting DM models are being constrained using only data that is insensitive to non-linear corrections.

In constraining the DM-b model, we consider a conservative data combination that does not include CMB lensing data, i.e. one that is not appreciably affected by non-linear corrections. We find 10-20\% improvement in the constraints on the DM-b scattering amplitude, $\sigma_{\mathrm{DM-b}}$, when combining SPT-3G primary CMB data with {\it{Planck}} and BAO data. Moreover, we confirm previous findings~\cite{Li:2022mdj} that adding ACT DR4~\cite{ACT_DR4} data to the previous ones relaxes the constraints on $\sigma_{\mathrm{DM-b}}$. Since there is a strong degeneracy between the spectral index of curvature perturbations, $n_s$, and $\sigma_{\mathrm{DM-b}}$ (see Figure~\ref{fig:dm_baryon_corner}), this result highlights the possibility of systematic biases affecting  $\Lambda$CDM parameters to alter constraints on extended models due to such degeneracies. In addition, using the Fisher formalism, we forecast up to 50\% improvement in constraints on $\sigma_{\mathrm{DM-b}}$ when adding lensing data from SPT-3G. This is a motivation to estimate non-linear corrections for such models, so that lensing data on small scales can be used to improve their constraints.

On the other hand, for the DM-DR case, we present constraints for the two cases $n_\mathrm{DM-DR}\in\{2,4\}$, using both parametrizations described in Section~\ref{subsec:DMDR}, and for data combination with and without lensing information. Our first finding is that for parametrization \texttt{param1}, which uses $\xi$ and $\log_{10}(a_\mathrm{dark})$ as model parameters, one can get meaningful constraints only when lensing data (from SPT-3G and ACT) is included. 
We remind the reader that our constraints derived using lensing data are to be taken with caution, as non-linear corrections are not included in the theoretical modeling.
However, for parametrization \texttt{param2}, in which $a_\mathrm{dark}\xi^4$ and $\Delta N_\mathrm{eff}$ are used,  even when lensing data is included one cannot extract fruitful constraints at 95\% CL on DM-DR parameters. Better constraints are obtained only when Lyman-$\alpha$ forest flux data is added, as done in~\cite{Archidiacono:2019wdp}. However, such data can be used only after accounting for non-linear corrections, which reinforces the key takeaway of this article.

\section*{Acknowledgments}
We thank Gil Holder and the South Pole Telescope (SPT) collaboration for helpful discussions and feedback.
We thank Srinivasan Raghunathan for providing post-ILC noise residuals for SPT-3G surveys used in our Fisher forecasts.
This work was supported by the Center for AstroPhysical Surveys (CAPS) at the National Center for Supercomputing Applications (NCSA), University of Illinois Urbana-Champaign. This work made use of the Illinois Campus Cluster, a computing resource that is operated by the Illinois Campus Cluster Program (ICCP) in conjunction with the National Center for Supercomputing Applications (NCSA) and which is supported by funds from the University of Illinois at Urbana-Champaign. This project has also received funding from the European Research Council (ERC) under the European Union’s Horizon 2020 research and innovation program (grant agreement No 101001897). Part of this work also made use of the Infinity Cluster hosted by IAP.

\bibliographystyle{utphys}
\bibliography{idm}

\end{document}